\documentclass[aps,prx,twocolumn,superscriptaddress,showpacs,longbibliography]{revtex4-1}
\usepackage{amsmath}
\usepackage{graphicx}
\usepackage{amsfonts}
\usepackage{verbatim}
\usepackage{amssymb}
\usepackage{color}
\usepackage{dsfont}
\usepackage{hyperref}
\usepackage{physics}
\hypersetup{colorlinks = true, urlcolor = blue, linkcolor = blue, citecolor = blue}

\newcommand{\su}{\uparrow} 
\newcommand{\sd}{\downarrow} 
\newcommand{\bpm}{\begin{pmatrix}}
\newcommand{\epm}{\end{pmatrix}}

\newcommand{\nn}{\nonumber \\} 

\newcommand{\dg}{^{\dagger}}

\begin{document}

\title{Enhancing the excitation gap of a quantum-dot-based Kitaev chain}

\author{Chun-Xiao Liu}\email{Corresponding author: chunxiaoliu62@gmail.com}
\affiliation{QuTech and Kavli Institute of Nanoscience, Delft University of Technology, Delft 2600 GA, The Netherlands}

\author{A. Mert Bozkurt}
\affiliation{QuTech and Kavli Institute of Nanoscience, Delft University of Technology, Delft 2600 GA, The Netherlands}

\author{Francesco Zatelli}
\affiliation{QuTech and Kavli Institute of Nanoscience, Delft University of Technology, Delft 2600 GA, The Netherlands}

\author{Sebastiaan L.D. ten Haaf}
\affiliation{QuTech and Kavli Institute of Nanoscience, Delft University of Technology, Delft 2600 GA, The Netherlands}

\author{Tom Dvir}
\affiliation{QuTech and Kavli Institute of Nanoscience, Delft University of Technology, Delft 2600 GA, The Netherlands}

\author{Michael Wimmer}
\affiliation{QuTech and Kavli Institute of Nanoscience, Delft University of Technology, Delft 2600 GA, The Netherlands}

\date{\today}

\begin{abstract}
Connecting double quantum dots via a semiconductor-superconductor hybrid segment offers a platform for creating a two-site Kitaev chain that hosts a pair of ``poor man's Majoranas'' at a finely tuned sweet spot. 
However, the effective couplings, which are mediated by Andreev bound states in the hybrid, are generally weak in the tunneling regime.
As a consequence, the excitation gap is limited in size, presenting a formidable challenge for using this platform to demonstrate non-Abelian statistics of Majoranas and realizing error-resilient topological quantum computing.
In this work, we systematically study the effects of increasing the coupling between the dot and the hybrid segment. 
In particular, the proximity effect transforms the dot orbitals into Yu-Shiba-Rusinov states, forming a new spinless fermion basis for a Kitaev chain, and we derive a theory for their effective coupling.
As the coupling strength between the dots and the hybrid segment increases, we find a significant enhancement of the excitation gap and reduced sensitivity to local perturbations.
Although the hybridization of the Majorana wave function with the central Andreev bound states increases strongly with increasing coupling, the overlap of Majorana modes on the outer dots remains small, which is a prerequisite for potential qubit experiments.
We discuss how the strong-coupling regime shows in experimentally accessible quantities, such as the local and non-local conductance, and provide a protocol for tuning a double-dot system into a sweet spot with a large excitation gap.
\end{abstract}

\maketitle

\section{Introduction}
The Kitaev chain is a toy model of topological superconductivity that consists of one-dimensional spinless fermions with $p$-wave pairing potential~\cite{Kitaev2001Unpaired}. 
In the topological phase, the endpoints host a pair of Majorana zero modes~\cite{Alicea2012New, Leijnse2012Introduction, Beenakker2013Search, Stanescu2013Majorana, Jiang2013Non, Elliott2015Colloquium, Sato2016Majorana, Sato2017Topological, Aguado2017Majorana, Lutchyn2018Majorana, Zhang2019Next, Frolov2020Topological}, which obey non-Abelian statistics and are regarded as the building block of topological quantum computation~\cite{Nayak2008Non-Abelian, DasSarma2015Majorana}.
Such a Majorana qubit is predicted to be more immune to decoherence due to the quantum information being encoded nonlocally in space and further protected by an excitation gap above the computational subspace.

In solid-state physics, the Kitaev chain model can be simulated in a quantum dot array by utilizing the spin-polarized dot orbitals as spinless fermions, with the effective couplings mediated by superconductivity~\cite{Sau2012Realizing}.
Remarkably, even a chain consisting of only two quantum dots can exhibit fine-tuned, but still spatially separated Majorana modes at a sweet spot, colloquially called poor man's Majorana modes~\cite{Leijnse2012Parity}.
Recently, such a two-site Kitaev chain was experimentally realized in double quantum dots, and poor man's Majorana modes were identified via conductance spectroscopy at the sweet spot~\cite{Dvir2023Realization}.
In particular, the effective couplings, both normal and superconducting ones, are mediated by an Andreev bound state (ABS) in a hybrid segment connecting both quantum dots~\cite{Liu2022Tunable}, which allows for a deterministic fine-tuning of the relative amplitude by changing the ABS chemical potential via electrostatic gating~\cite{Wang2022Singlet, Wang2023Triplet, Bordin2023Tunable}. This effect was shown theoretically to be robust to Coulomb interactions in the dots as well as stronger coupling~\cite{Tsintzis2022Creating}.

Despite the experimental progress, state-of-the-art Kitaev chain devices are still constrained by a relatively small excitation gap ($ \sim 25~\mu$eV), which is much smaller than the induced gap of the ABS ($ \sim 150~\mu$eV) and the parent aluminum gap ($ \sim 230~\mu$eV)~\cite{Dvir2023Realization}.
In order to experimentally demonstrate the non-Abelian statistics of Majoranas and to obtain high-quality Majorana qubits~\cite{Liu2023Fusion, Boross2023Braiding, Tsintzis2023Roadmap}, a significant enhancement in the excitation gap is crucial.
This enhancement will allow for a more tolerant adiabatic limit condition $\sim \hbar/E^{-1}_{\text{gap}}$~\cite{Aasen2016Milestones, Knapp2016Nature, Nag2019Diabatic, Bauer2018Dynamics} and suppress the detrimental thermal effects $\sim e^{-E_{\text{gap}}/k_BT}$~\cite{Karzig2021Quasiparticle}.

In this work, we use the three-site model~\cite{Liu2022Tunable, Tsintzis2022Creating, Dominguez2016Quantum} to systematically study enhancing the energy gap by increasing the dot-hybrid coupling strength, achievable in experiment by lowering the tunnel barrier height.
As a result of the proximity effect from the hybrid, the orbitals in the quantum dots undergo a transformation into Yu-Shiba-Rusinov (YSR) states~\cite{Yu1965Bound, Shiba1968Classical, Rusinov1969Theory}.
These states then constitute the new spinless fermion basis for the emulated Kitaev chain. 
Thus, the concepts of elastic cotunneling and crossed Andreev reflection in the weak coupling regime have to be generalized.
Most importantly, we show that poor man's Majorana zero modes can survive in this strong coupling regime, featuring a significantly enhanced excitation gap.
The properties of the resulting states are different from those in the weak coupling regime, showing both in wavefunction profiles and conductance properties, while maintaining their Majorana character.

\begin{figure}
\centering
{\includegraphics[width = \linewidth]{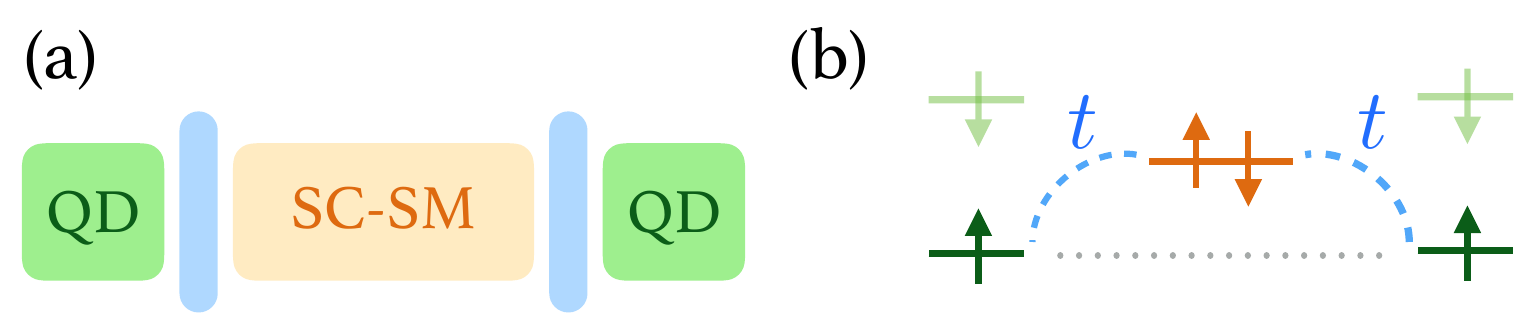}}
\caption{(a) Schematic of a two-site Kitaev chain device. Two separated quantum dots (green) are connected by a hybrid segment (orange) in the middle, with the strength of the dot-hybrid coupling being controlled by the tunnel gates (blue). (b) Schematic of the dot orbitals and Andreev bound states introduced in the model Hamiltonian.}
\label{fig:schematic}
\end{figure}

\section{Model and Hamiltonian}
A two-site Kitaev chain device consists of two separated quantum dots connected by a hybrid segment [see Fig.~\ref{fig:schematic}(a)].
The system Hamiltonian is~\cite{Liu2022Tunable, Tsintzis2022Creating, Dominguez2016Quantum}
\begin{align}
&H = H_D + H_S + H_T, \nn
&H_D = \sum_{a=L,R} (\varepsilon_{Da} + E_{ZDa}) n_{Da\su}
+ (\varepsilon_{Da} - E_{ZDa}) n_{Da\sd} \nn
&\quad + U_{Da} n_{Da\su}n_{Da\sd}, \nn
&H_S = \varepsilon_A (n_{A\su} + n_{A\sd}) + \Delta_0 (c_{A\su} c_{A\sd} + c\dg_{A\sd} c\dg_{A\su} ), \nn
&H_T = \sum_{\sigma=\su,\sd} \Big( t_L c\dg_{A\sigma} c_{DL\sigma} + \sigma t_{Lso} c\dg_{A\overline{\sigma}} c_{DL\sigma} \nn
&\qquad + t_R c\dg_{DR\sigma} c_{A\sigma} + \sigma t_{Rso} c\dg_{DR\overline{\sigma}} c_{A\sigma} \Big) + h.c.,
\label{eq:Ham_three_site}
\end{align}
where $H_D$ is the Hamiltonian of the quantum dots, $n_{Da\sigma}=c\dg_{Da\sigma}c_{Da\sigma}$ is the electron occupancy number on dot $a$, $\varepsilon_{Da}$ is the orbital energy, $E_{ZDa}$ is the strength of the induced Zeeman energy, and $U_{Da}$ is the Coulomb repulsion strength.
$H_S$ describes the hybrid segment hosting a pair of ABSs in the low-energy approximation.
$\varepsilon_A$ is the normal-state energy, and $\Delta_0$ is the induced pairing gap.
While we assume no induced Zeeman energy in the ABS due to a strong renormalization effect at the hybrid interface~\cite{Stanescu2010Proximity, Antipov2018Effects}, the main conclusions remain valid for finite Zeeman energy as well.
$H_T$ is the tunnel coupling between dot and ABS, including both spin-conserving $\sim t$ and spin-flipping $\sim t_{so}$ processes.
In realistic devices, the amplitude of $t$ is a variable that can be controlled by tunnel barrier gates, while the ratio of $t_{so}/t$ is generally fixed and is determined by the strength of spin-orbit interaction.
In the rest of this work, we will choose $\Delta_0$ to be the natural unit. 
Unless stated otherwise, we set $E_{ZDa}=1.5~\Delta_0$, $U_{Da}=5~\Delta_0$, and $t_{aso}/t_a=0.3$ according to the recent experimental measurements on similar devices~\cite{Wang2022Singlet, Wang2023Triplet, Dvir2023Realization, Bordin2023Tunable}.
In addition, we numerically calculate the differential conductance using the rate-equation method~\cite{Tsintzis2022Creating}, where the lead tunneling rate is $\Gamma_a = 0.025 \Delta_0$, and temperature is $k_BT=0.02~\Delta_0$.

\begin{figure}
\centering
{\includegraphics[width = \linewidth]{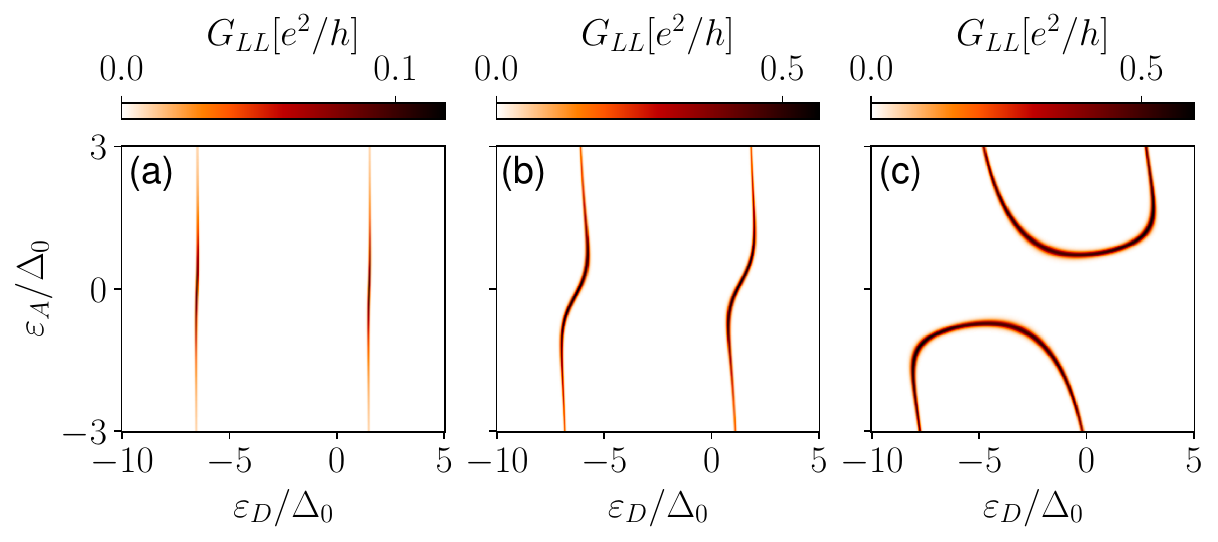}}
\caption{Local conductance spectroscopy for a dot-ABS pair. 
The tunneling strength is $t/\Delta_0=0.25, 1, 2$ in panels (a), (b) and (c), respectively.
}
\label{fig:dot_ABS}
\end{figure}

\section{quantum dot-Andreev bound state pair}

To assess the strength and to understand the effects of the dot-hybrid coupling, we first focus on the conductance spectroscopy of a single quantum dot-ABS pair.
Hence, for the discussions in this section, we temporarily remove the right dot in the model Hamiltonian in Eq.~\eqref{eq:Ham_three_site}.
Figure~\ref{fig:dot_ABS} shows the zero-bias conductance spectroscopy in the $(\varepsilon_D, \varepsilon_A)$ plane for $t/\Delta_0=0.25, 1$ and $2$, respectively.
As shown in Fig.~\ref{fig:dot_ABS}(a), in the weak coupling regime, the conductance resonances are two straight lines extending along $\varepsilon_A$, corresponding to the spin-up and down orbitals in the quantum dot.
In contrast, with a strong dot-hybrid coupling, the resonance lines become $S$-shaped curves [see Fig.~\ref{fig:dot_ABS}(b)], and the conductance magnitudes are increased by order of magnitude. 
For even stronger coupling, Fig.~\ref{fig:dot_ABS}(c), proximity from the ABS is so strong that the states are very different from spin-up and spin-down, as signified by the differently connected arcs in the conductance.

This qualitative behavior was already described in Ref.~\cite{Rasmussen2018YSR}. Here, we recover the same behavior in a simpler model and also consider the magnitude of the conductance. In particular, we can use second-order perturbation theory to qualitatively understand the physical mechanisms underlying these conductance features.

First, the dot-hybrid coupling renormalizes the dot orbital energy by $\delta \varepsilon_D n_{D}$ via cotunneling processes.
Up to the leading order of $t$ and $t_{so}$, this energy renormalization is
\begin{align}
    \delta \varepsilon_D = (t^2 + t^2_{so}) \frac{u^2 - v^2}{E_{A}} + O(t^4, t^4_{so}),
    \label{eq:dot_shift}
\end{align}
where $u^2=1-v^2=1/2 + \varepsilon_A/2E_A$ are the BCS coherence factors, and $E_A = \sqrt{\varepsilon^2_A + \Delta^2_0}$ is the ABS excitation energy.
Interestingly, due to destructive interference, the dot energy shifts positively (negatively) for $\varepsilon_A>0$ ($\varepsilon_A<0$), vanishes at $\varepsilon_A=0$, and decays as $\varepsilon^{-1}_A$ for large $\varepsilon_A$, well explaining the $S$-shaped conductance resonances shown in Fig.~\ref{fig:dot_ABS}(b).
Hence, the $S$-shaped feature is a clear sign of the proximity effect due to the ABS in the hybrid semiconductor-superconductor segment (rather than directly to the parent metallic superconductor), and the increase in bending is a direct signal of increasing coupling.

From Figs.~\ref{fig:dot_ABS}(a) and \ref{fig:dot_ABS}(b) we also observe that conductance is largest when the ABS is near resonance, i.e., $-\Delta_0 < \varepsilon_A < \Delta_0$, which can also be understood from perturbation theory.
The ABS will induce a pairing term on the quantum dot, i.e., $\Delta_{\text{ind}} c\dg_{D\su} c\dg_{D\sd} + h.c.$  via local Andreev reflection with
\begin{align}
\Delta_{\text{ind}} = (t^2 + t^2_{so} ) \frac{2uv}{E_{A}} + O(t^4, t^4_{so}).
\label{eq:dot_induced}
\end{align}
The induced gap size is prominent within $-\Delta_0 < \varepsilon_A < \Delta_0$ and decays as $\varepsilon^{-2}_A$ outside.
As a result, the local Andreev conductance is significantly enhanced when ABS is near resonance and vanishes when off-resonance.
Thus, based on the shape of the resonance lines as well as on the enhancement of the Andreev conductance, one can estimate the strength of the dot-hybrid coupling in an actual device.
More importantly, especially in the strong coupling regime, the induced superconducting effect would transform the dot orbitals into YSR states~\cite{Satori1992Numerical, Bauer2007Spectral, Zitko2011Effects, Hatter2015Magnetic, Rasmussen2018YSR, Yao2014Phase, scherubl2020, scherubl2022}.
These states then establish the new spinless fermion basis for the emulated Kitaev chain.

\begin{figure}
\centering
{\includegraphics[width = \linewidth]{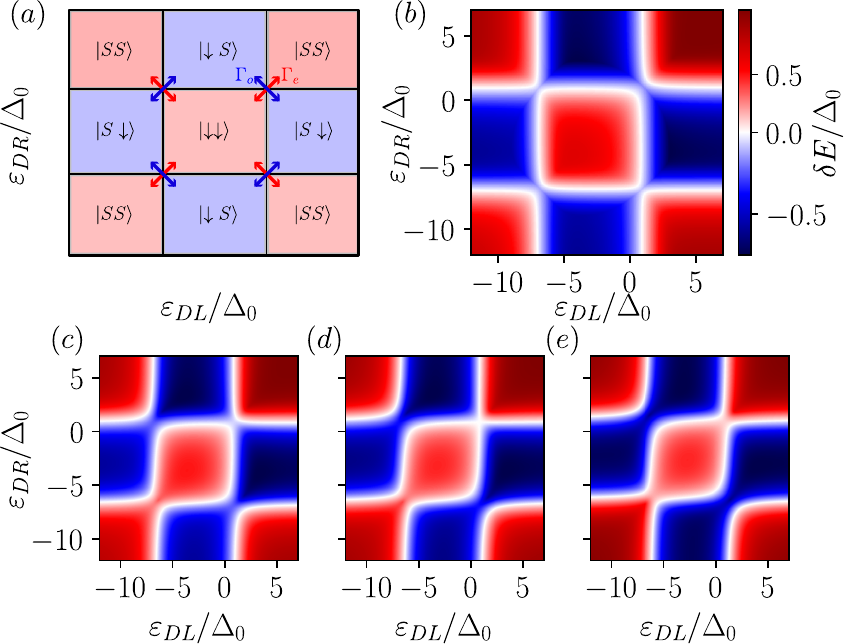}}
\caption{Upper panels: (a) Schematic description of the charge stability diagram. $\Gamma_e$ and $\Gamma_o$ couples even and odd states, respectively. (b) Charge stability diagram with $t_{so}=0$ and $\varepsilon_A = -1.5\Delta_0$. In this case, only spin conserving processes are allowed, resulting in a ring-like pattern in the charge stability diagram. Lower panels: Changing $\varepsilon_A$ alters the charge stability diagram with the upper-right corner is (c) $\Gamma_o$ dominated with $\varepsilon_A = \varepsilon_A^* - 0.2\Delta_0$, (d) at the sweet spot $\varepsilon_A = \varepsilon_A^*$ and (e) $\Gamma_e$ dominated $\varepsilon_A = \varepsilon_A^* + 0.2\Delta_0$ with $\varepsilon_A^*\approx-0.269\Delta_0$ being the sweet spot value. In panels (b-e), we have $t=\Delta_0$.
}
\label{fig:CSD}
\end{figure}

\section{Coupled YSR states}
We now turn to the case of two quantum dots coupled via an ABS, and develop an effective theory for two coupled YSR states.
By assuming that the ABS in the hybrid remain gapped, we can integrate it out and obtain the effective coupling 
\begin{align}
H^{\text{eff}}_{\text{coupling}} =\sum_{\sigma, \eta=\su, \sd} \left( t_{\sigma \eta} c\dg_{DL \sigma} c_{DR \eta} + \Delta_{\sigma \eta} c\dg_{DL \sigma} c\dg_{DR \eta} \right) + h.c.,
\label{eq:H_eff_couple}
\end{align} 
where $t_{\sigma \eta}$ and $\Delta_{\sigma \eta}$ are the elastic cotunneling (ECT) and crossed Andreev reflection (CAR) amplitudes between electron or hole excitations in the two dots. 
These couplings are tunable by changing the energy of the middle dot, $\varepsilon_A$ \cite{Liu2022Tunable}.
Note that the problem of coupling two quantum dots via ECT and CAR, and in the presence of local Andreev reflection giving rise to a proximity effect in the dots has been studied extensively before~\cite{eldridge2010, su2017, scherubl2019Transport, kurtossy2021}, predominantly at zero magnetic field.
In contrast, we focus on the case of a significant Zeeman splitting in the outer quantum dots, such that the ground state of both dots occupied by a single electron is a triplet state. 

For unproximitized quantum dots, Eq.~\eqref{eq:H_eff_couple} plus $H_D$ in Eq.~\eqref{eq:Ham_three_site} indeed resembles the Hamiltonian of a two-site Kitaev chain~\cite{Kitaev2001Unpaired}.
However, since YSR states are a superposition of electron and hole components, the effective couplings of ECT and CAR have to be generalized.
In particular, for a single proximitized quantum dot with finite Zeeman splitting, the ground states in the even- and odd-parity subspaces are a spin singlet and a spin-down state, respectively,
\begin{align}
 \ket{S} = u | 00 \rangle - v | 11 \rangle, \quad \ket{\sd} = \ket{01}
 \label{eq:singlet}
\end{align}
where $u^2 = 1 - v^2 = 1/2 + \xi/2E_0$ are the BCS coherence factors, with $\xi = \varepsilon + U/2$ and $E_0 = \sqrt{\xi^2 + \Delta^2_{\text{ind}}}$, and $\ket{n_{\su} n_{\sd}}$ is a state in the occupancy representation. 
Consequently, we define YSR state as $\ket{\sd} = f\dg_{\text{YSR}}  \ket{S}$, with an excitation energy $\delta \varepsilon = E_{\sd} - E_{S}$.
When coupling two YSR states via the ABS, the effective Hamiltonian becomes
\begin{align}
H_{\text{eff}} = \sum_{a=L,R} \delta \varepsilon_a f\dg_a f_a + \Gamma_o f\dg_L f_R +  \Gamma_e f\dg_L f\dg_R + h.c.,
\end{align}
where $f_a$ denotes the YSR state in dot-$a$, with $\delta \varepsilon_a$ being the excitation energy.
This also takes the form of a Kitaev Hamiltonian, but now in the basis of YSR states.
Crucially, $\Gamma_{o/e}$ represent the generalized effective couplings between YSR states.
The odd-parity coupling is 
\begin{align}
\Gamma_o &= \bra{S\sd} H^{\text{eff}}_{\text{coupling}}  \ket{\sd S}, \nn
&= -t_{\su \su} v_L v_R + t_{\sd \sd} u_L u_R + \Delta_{\su \sd} v_L u_R -\Delta_{\sd \su} u_L v_R,
\label{eq:gamma_o}
\end{align}
where $\ket{\sd S}, \ket{S \sd}$ are the tensor states with total parity odd, and $u_a, v_a$ are the BCS factors defined in Eq.~\eqref{eq:singlet}.
Note that $\Gamma_o$ is a linear combination of equal-spin ECT and opposite-spin CAR, which are all spin-conserving processes.
On the other hand, the even-parity coupling is
\begin{align}
\Gamma_e &= \bra{S S} H^{\text{eff}}_{\text{coupling}}  \ket{\sd \sd}, \nn
&=  -\Delta_{\su \su} v_L v_R + \Delta_{\sd \sd} u_L u_R  + t_{\su \sd} v_L u_R -t_{\sd \su} u_L v_R,
\label{eq:gamma_e}
\end{align}
which couples states with total spin zero and one.
In particular, a finite $\Gamma_e$ requires a physical mechanism to break spin conservation, e.g., spin-orbit interaction.

Figure~\ref{fig:CSD}(a) shows a schematic of the charge stability diagram as a function of the quantum dot energies.
Blue and red squares indicate whether the ground state of two uncoupled dots is odd or even, respectively. 
Note that the singlet ground state in the dots does depend on the dot energies through the values of $u_a$ and $v_a$.
For example, in the upper-right corner $u_{L/R} > v_{L/R}$, corresponding to each dot predominantly empty, whereas the lower-left corner features $u_{L/R} < v_{L/R}$, corresponding to each dot predominantly doubly occupied.
The arrows represent the interactions $\Gamma_\text{o/e}$, and the relative strength of these couplings will
determine the ground state close to the four corners in the charge stability diagram where each dot exhibits a degeneracy without interactions. 
Additionally, we point out that the role of CAR and ECT interchange for different corners of the charge stability diagram.

In the absence of spin-orbit interaction, $\Gamma_\text{e}=0$, and in general $\Gamma_\text{o}\neq 0$.
Hence, at the four corners, the energy of the odd ground state is lowered compared to the even ones.
This can also be observed in a simulation of the full three-dot Hamiltonian in Eq.~\eqref{eq:Ham_three_site} as shown in Fig.~\ref{fig:CSD}(b), where we find a disconnected even island in the center of the charge stability diagram. 
Note that such a behavior is only possible for a finite Zeeman splitting, and as such qualitatively different from the charge stability diagrams in Ref.~\cite{scherubl2019Transport}.

For a system with finite spin-orbit coupling, in general also $\Gamma_\text{e}\neq 0$, and the respective values will depend on details of the system (such as the energy $\varepsilon_A$ of the middle dot that can be used to tune ECT and CAR).
In particular, it is now in general possible to change the relative strength of $\Gamma_\text{e/o}$. 
This shows as a change in connectivity in the charge stability diagram, with a guaranteed sweet spot $\Gamma_\text{e} = \Gamma_\text{o}$ in between.
We show this behavior in Fig.~\ref{fig:CSD}(c-e) on the example of the upper-right corner as we vary $\varepsilon_a$, transitioning from a regime dominated by $\Gamma_o$ in Fig.~\ref{fig:CSD}(c) to one dominated by $\Gamma_e$ in Fig.~\ref{fig:CSD}(e).
When $\Gamma_e = \Gamma_o$, a cross emerges in the phase diagram as a signature of sweet spot, as shown in Fig.~\ref{fig:CSD}(d).

In the limit of large Coulomb interaction $U$ on the outer dots, either $u_\text{L/R}\approx 1$ or $v_\text{L/R}\approx 1$, and $\Gamma_\text{e/o}$ will be dominated by a single ECT or CAR term, as evident from Eqs.~\eqref{eq:gamma_o} and \eqref{eq:gamma_e}. 

More generally however, in the limit of vanishing Zeeman splitting in the middle dot, CAR and ECT coupling are constrained by
$t_{\su \su} = t_{\sd \sd}, t_{\su \sd} = -t_{\sd \su}, \Delta_{\su \su} = \Delta_{\sd \sd}, \Delta_{\su \sd} = -\Delta_{\sd \su}$,
due to time reversal symmetry, such that $\Gamma_{o/e}$ can be further simplified as
\begin{align}
&\Gamma_{o} =  t_{\su \su} \cos( \beta_L + \beta_R ) + \Delta_{\su \sd} \sin( \beta_L + \beta_R ), \nn
&\Gamma_{e} =   \Delta_{\su \su} \cos( \beta_L + \beta_R ) + t_{\su \sd} \sin( \beta_L + \beta_R ),
\end{align}
where $0 \leq \beta_a \leq \pi/2$ is a parameter to characterize the BCS factors by $u_a = \cos \beta_a, v_a = \sin \beta_a$.
For two dots with a similar level of proximity effect, the diagonal corners in Fig.~\ref{fig:CSD} will have $\beta_L = \beta + \delta/2, \beta_R = \beta - \delta/2$, with $\delta \ll 1$ characterizing a weak asymmetry.
As a result, the odd- and even-parity couplings reduce to
\begin{align}
&\Gamma_{o}  \approx t_{\su \su} \cos(2\beta) + \Delta_{\su \sd} \sin(2\beta), \nn
&\Gamma_{e}  \approx \Delta_{\su \su} \cos(2\beta) + t_{\su \sd} \sin(2\beta).
\end{align}
This indicates that as the proximity effect increases, the initially purely equal-spin ECT/CAR coupling ratio $\Gamma_{o}/\Gamma_e$ gains a finite opposite-spin CAR/ECT component.
In contrast, around the off-diagonal corners, we have $\beta_L = \beta + \delta/2,  \pi/2 - \beta_R = \beta - \delta/2$, yielding
\begin{align}\label{eqn:offdiagonal_gammas}
&\Gamma_{o} \approx \Delta_{\su \sd} \cos(\delta) - t_{\uparrow \uparrow} \sin(\delta), \nn
& \Gamma_{e} \approx t_{\su \sd} \cos(\delta) - \Delta_{\su \su} \sin (\delta).
\end{align}
Interestingly, despite the proximity effect, $\Gamma_{o/e}$ are equal to the only ECT or CAR just as in the unproximitized regime. 
Only an asymmetry in the proximity effect leads to a mixing of CAR and ECT-type couplings.
For a detailed investigation of how $\Gamma_e$ and $\Gamma_o$ behave in different corners of the charge stability diagram, we refer to the Supplementary material.

\section{Poor man's Majorana}

\begin{figure}[t]
\centering
{\includegraphics[width = \linewidth]{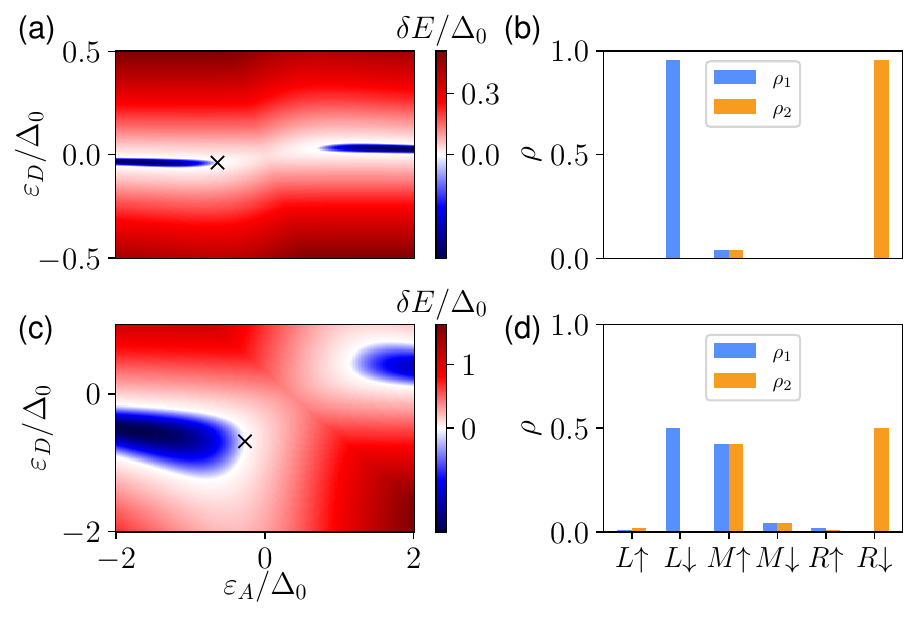}}
\caption{
Upper panels: (a) Phase diagram and (b) Majorana wavefunction profiles in the weak dot-ABS coupling regime ($t/\Delta_0=0.25$).
Lower panels: same physical quantities calculated in the strong coupling regime ($t/\Delta_0=1$).
In panels (b) and (d), bars in blue and orange denote two Majoranas profiles defined in Eq.~\eqref{eq:rho}.
}
\label{fig:phase_diagram}
\end{figure}

We now focus on the properties of the poor man's Majoranas that appear at the sweet spot in the full dot-hybrid-dot system.
Without loss of generality, we assume that the left and right dots have the same set of physical parameters, e.g., $\varepsilon_{DL} = \varepsilon_{DR} =\varepsilon_{D}, E_{ZL}=E_{ZR}=1.5\Delta_0, U_L=U_R=5\Delta_0, t_L=t_R=t$ and $t_{Lso}=t_{Rso}=0.3t$.
To simplify, we introduce a shift in dot energy $\varepsilon_{D} \to \varepsilon_{D} - E_{ZD}$ to set the zero energy of the spin-down orbital at $\varepsilon_{D}=0$.
Figure~\ref{fig:phase_diagram}(a) shows the phase diagram in the $(\varepsilon_A, \varepsilon_{D})$ plane for weakly coupled quantum dots ($t/\Delta_0=0.25$), with $\delta E = E^{\text{odd}}_{\text{gs}} - E^{\text{even}}_{\text{gs}}$ being the energy difference of the ground states in the opposite fermion parity subspace.
The white-colored curves ($\delta E=0$) represent the ground state degeneracy, with the tip of the curves (marked by a black cross sign) indicating the sweet spot~\cite{Tsintzis2022Creating}.
At this sweet spot the effective normal and superconducting couplings of the two dots become equal in strength.
The wavefunction profiles plotted in Fig.~\ref{fig:phase_diagram}(b) further demonstrate that the Majorana zero modes are well localized at the two dots, respectively, with only a negligible amount of overlap in the middle ABS.
Here, the Majorana wavefunction densities are defined for site and spin
\begin{align}
& \rho_{1a \sigma} = | \langle o | ( c_{a \sigma} + c\dg_{a \sigma} ) | e \rangle  |^2, \nn
& \rho_{2a \sigma} = | \langle o | ( c_{a \sigma} - c\dg_{a \sigma} ) | e \rangle  |^2,
\label{eq:rho}
\end{align}
where $a=L,M, R$, $\sigma=\su, \sd$, and $|e \rangle$, $|o \rangle$ denote the even- and odd-parity ground state.

Comparatively, the lower panels in Fig.~\ref{fig:phase_diagram} show the results obtained in the strong dot-ABS coupling regime with $t/\Delta_0=1$.
Around the sweet spot marked by the black cross sign, the ground state degeneracy line now becomes much broader and straighter compared to the weak coupling regime, indicating a significantly enhanced energy gap and robustness against dot chemical potential fluctuations.
In Fig.~\ref{fig:phase_diagram}(d), the plotted Majorana wavefunctions show strong leakage into the middle ABS and small leakage to the opposite normal dot with opposite spin.
We emphasize that the wavefunction overlap in the ABS, which is a virtual state, is not detrimental, and that the reduced density on the normal dots will only reduce the visibility of the Majorana from the external detecting system.

\begin{figure}[t]
\centering
{\includegraphics[width = \linewidth]{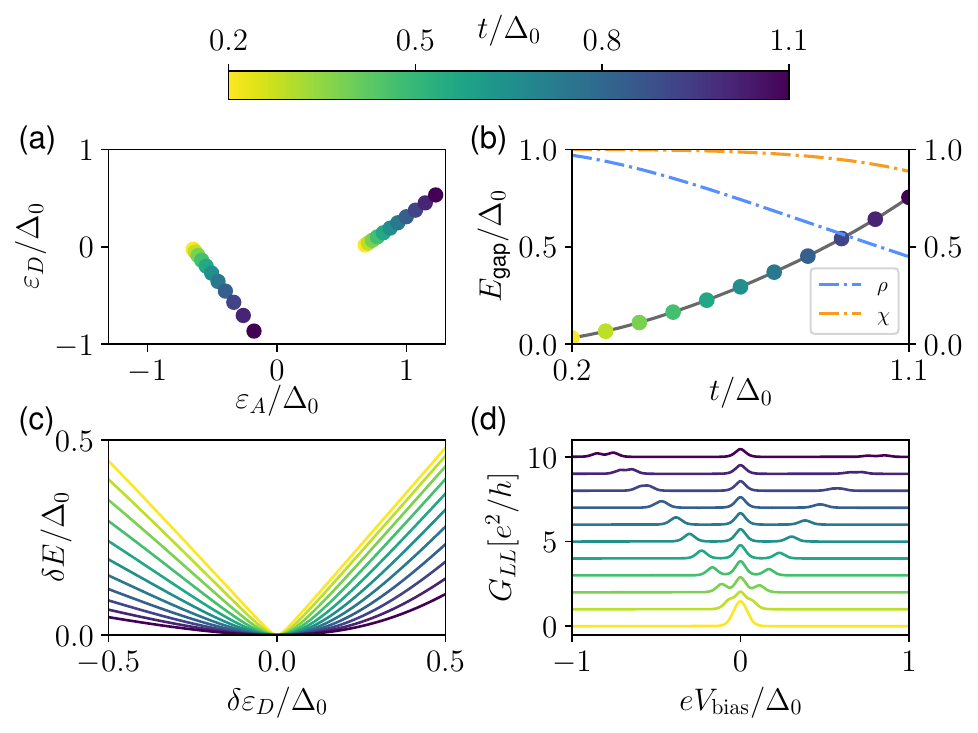}}
\caption{The evolution of the relevant quantities of poor man's Majorana as a function of the coupling strength $t$. (a) shows the evolution of the position of sweet spot in the phase diagram as a function of $t$. In (b), we depict the evolution of the excitation gap at the sweet spot and the Majorana density $\rho$ and polarization $\chi$ on the left dot. (c) depicts the ground state degeneracy splitting when both dots chemical potentials are detuned away from the sweet spot. As $t$ increases, the amount of splitting diminishes. 
(d) Waterfall plots of local conductance profile at the sweet spot for each $t$. Even though the main peak height decreases as $t$ increases, the side peaks that signify the excitation gap appear at a larger bias and furthermore split.}
\label{fig:trend}
\end{figure}

To gain a better understanding of the effect of strong coupling, we now investigate the continuous evolution of the sweet spot and the properties of Majoranas as a function of $t$.
To that end, we define the following quantities: excitation gap ($E_{\text{gap}}$), Majorana localization ($\rho$) and polarization ($\chi$) \cite{Tsintzis2022Creating} as below
\begin{align}
& E_{\text{gap}} = \text{min}( \delta E^1_{oe},  \delta E^1_{eo} ), \nn
& \rho = \rho_{1L}=\rho_{1L\su} + \rho_{1L\sd}, \nn
& \chi = (\rho_{1L} - \rho_{2L })/(\rho_{1L} + \rho_{2L}).
\end{align}
Here, $E_{\text{gap}}$ represents the excitation gap above the poor man's Majorana zero modes, where $\delta E^1_{oe}$ and $\delta E^1_{eo}$ denote the energy differences between the ground state in one parity sector and the first excited state in the opposite one.
The Majorana localization $\rho$ and Majorana polarization $\chi$, both of which are defined on the outer quantum dot, characterize the localization and overlap of the Majorana wavefunction on the normal dot, respectively.

Figure~\ref{fig:trend}(a) shows the evolution of the positions of the sweet spots in the $(\varepsilon_A, \varepsilon_{D})$ plane.
As the coupling strength $t$ increases, the effect of dot energy renormalization predicted by Eq.~\eqref{eq:dot_shift} becomes more pronounced, making $\varepsilon_{D}$ deviate from the value of $\varepsilon_{D}=0$ in the weak coupling regime.
At the same time, the sweet-spot values of $\varepsilon_{A}$ shift towards a more positive value, indicating an induced Zeeman energy in ABS~\cite{Liu2022Tunable}, which comes from the inverse proximity effect from the quantum dot.
One crucial aspect of the strong coupling regime is that, with increasing $t$, the excitation gap is enhanced significantly in a nearly quadratic manner [see Fig.~\ref{fig:trend}(b)].
The excitation gap reaches as high as $E_\text{gap} \sim 0.7 \Delta_0$ for a dot-hybrid coupling $t/\Delta_0 = 1.1$.
Moreover, we observe that the degree of protection against detuning of both quantum dots away from the sweet spot increases with the growing $t$, as evidenced by the diminishing curvature of the quadratic splitting of ground state degeneracy in Fig.~\ref{fig:trend}(c).
On the other hand, the Majorana localization $\rho$ is largely reduced due to increased wavefunction leakage into the middle ABS when the tunnel barrier is lowered.
Yet, although the middle dot hybridizes strongly with the Majoranas in the outer dots, the Majorana polarization $\chi$ decreases much slower.
Hence, even for strong coupling, the overlap of Majoranas on the outer dots remains small which will be beneficial for future qubit experiments.

We observe these changes manifesting in the local conductance profile calculated at the sweet spot [see Fig.~\ref{fig:trend}(d)].
Specifically, as $t$ increases, the height of the Majorana-induced zero-bias conductance peaks decreases due to the reduction in Majorana density at the outer dot, which in turn reduces its effective coupling strength with the normal lead.
In addition, the voltage bias values where the side peaks appear, indicating the magnitude of the excitation gap, increases with $t$ and a single side peak begins to split into two at larger $t$ values, consistent with our findings in Fig.~\ref{fig:trend}(b).

\section{Signatures of strong coupling in nonlocal transport}

Nonlocal transport is a useful tool for probing hybrid systems, as the nonlocal conductances $G_{LR}$ and $G_{RL}$ can measure the BCS charge of states~\cite{danon2020Nonlocal} in non-interacting systems. 
For example, in Ref.~\cite{Dvir2023Realization}, the nonlocal conductance was used to confirm the chargeless Majorana character of the zero-energy state.

\begin{figure}[t]
\centering
{\includegraphics[width = \linewidth]{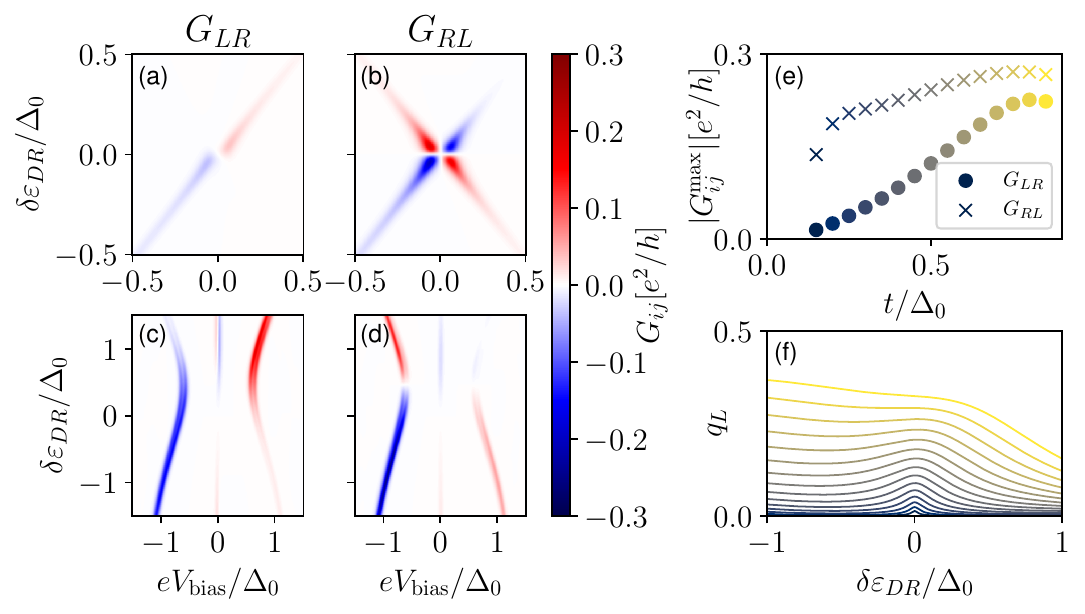}}
\caption{Nonlocal conductances in (a-b) the weak coupling ($t=0.25\Delta_0$) and (c-d) strong coupling regimes ($t=\Delta_0$): In both scenarios, we detune the right dot chemical potential from the sweet spot by $\delta\varepsilon_{DR}$. In the strong coupling regime, we observe an increased gap, while the maximum conductance signal for $G_{LR}$ strengthens, whereas $G_{RL}$ remains relatively stable. Panel (e) illustrates how the maximum conductance signal strength evolves with varying $t$, indicated by the color of the circles. As we increase the dot-hybrid coupling $t$, $G_{LR}$ shows significant enhancement, while $G_{RL}$ remains mostly constant. This increase in $G_{LR}$ is attributed to the increased BCS charge of the excited state in the left quantum dot with increasing $t$, leading to an increased conductance signal magnitude.}
\label{fig:nonlocal_transport}
\end{figure}

We now discuss how the visibility of the first excited state in the non-local conductance $G_{LR}$ is a qualitative indicator of the strong coupling regime.
In Fig.~\ref{fig:nonlocal_transport}, we show the nonlocal conductances in both weak coupling ($t=0.25~\Delta_0$) and strong coupling ($t=\Delta_0$) regimes, as a function of the applied voltage bias $V_\textrm{bias}$ and chemical potential detuning away from the sweet spot $\delta\varepsilon_{DR}$.
In the weak coupling regime, shown in Fig.~\ref{fig:nonlocal_transport}(a-b), the conductance signal strength for $G_{LR}$ is significantly lower than its counterpart $G_{RL}$.
This behavior is in line with the results for a spinless Kitaev chain model presented in Ref.~\cite{Dvir2023Realization}.
There, the chargeless nature of the excited state leads to a vanishing $G_{LR}$ signal.
In contrast, as we increase the dot-hybrid coupling, this effective picture breaks down and the nonlocal conductances $G_{LR}$ and $G_{RL}$ become comparable in strength, as demonstrated in Fig.~\ref{fig:nonlocal_transport}(c-d).
Notably, $G_{RL}$ for the first excited state changes sign as a function of detuning $\delta\varepsilon_{DR}$ in the right dot, whereas $G_{LR}$ does not.

In Fig.~\ref{fig:nonlocal_transport}(e), we track the evolution of maximum nonlocal conductance signal strength with varying $t$.
We observe that the maximum conductance for $G_{LR}$ increases much faster with increasing $t$ compared to $G_{RL}$. 
We can attribute the increase in $G_{LR}$ due to the increase in the BCS charge of the excited state in the left quantum dot with increasing $t$, as shown in Fig.~\ref{fig:nonlocal_transport}(f)~\cite{danon2020Nonlocal}.

\section{Discussion}
In this work, we have studied a dot-hybrid-dot system in the strong dot-hybrid coupling regime ($t \sim \Delta_0$) for implementing a two-site Kitaev chain.
Due to the proximity effect from the ABS, the dot orbitals undergo a transformation into YSR states, which constitute the new spinless fermion basis for the effective Kitaev chain, and we have studied their coupling as mediated by the middle dot.
Importantly, poor man's Majorana zero modes persist in this strong coupling regime, and now possess a significantly enhanced excitation gap.
On the other hand, there is an upper bound for the dot-hybrid coupling strength.
As shown in Fig.~\ref{fig:dot_ABS}(c), an excessive strong coupling $t \sim 2E_{ZD}$ leads to the hybridization of dot orbitals with opposing spins, gapping out the zero-energy YSR states.
That is, a sufficiently strong Zeeman field ($E_{ZD} \gg t$) is always a prerequisite for obtaining an effectively spinless Kitaev chain model in a spinful physical system~\cite{Pan2023Majorana, tenHaaf2023}.

Additionally, our theoretical work provides a practical recipe for implementing a two-site Kitaev chain with enhanced excitation gap.
Specifically, one can reach the desirable dot-hybrid coupling regime by observing the conductance spectroscopy of a single quantum dot-ABS while shifting the other dot off-resonance, e.g., Fig.~\ref{fig:dot_ABS}(b).
Performing this procedure for both left and right pairs, the sweet spot can be further fine-tuned by changing the chemical potential of the ABS to find a crossing in the charge stability diagram [see Fig.~\ref{fig:CSD}(c)-(e)] and a robust zero-bias peak in the conductance spectroscopy [see Fig.~\ref{fig:trend}(d)].
Indeed, a parallel experimental work~\cite{Zatelli2023Robust} has achieved an energy gap of approximately $\sim 75~\mu$eV using the aforementioned procedure, verifying the theoretical predictions proposed in this work.
We thus expect that our work provides  practical guidelines and physical insights for realizing Kitaev chains with an enhanced gap, serving as the central platform for future researches on Majorana quasiparticles and non-Abelian statistics.

\section*{Code availability}

The code used to generate the figures is available on Zenodo~\cite{ZenodoCode}.

\begin{acknowledgements}
This work was supported by a subsidy for top consortia for knowledge and innovation (TKl toeslag), by the Dutch Organization for Scientific Research (NWO), by by the European Union’s Horizon 2020 research and innovation programme FET-Open Grant No. 828948 (AndQC), and by Microsoft Corporation.

\emph{Author contributions.}---C.-X.L. designed the project with input from F.Z., S.L.D.t.H., T.D. and M.W.
C.-X.L. and A.M.B. performed the calculations and generated the figures.
All authors contributed to the discussions and interpretations of the results.
C.-X.L. and M.W. supervised the project.
C.-X.L., A.M.B., and M.W. wrote the manuscript with input from F.Z., S.L.D.t.H., and T.D.

\end{acknowledgements}

\bibliography{references_CXL.bib, strong_coupling}

\onecolumngrid
\newpage
\vspace{1cm}
\begin{center}
{\bf\large Supplemental Materials for ``Enhancing the excitation gap of a quantum-dot-based Kitaev chain"}
\end{center}
\vspace{0.5cm}

\setcounter{secnumdepth}{3}
\setcounter{equation}{0}
\setcounter{figure}{0}
\renewcommand{\theequation}{S-\arabic{equation}}
\renewcommand{\thefigure}{S\arabic{figure}}
\renewcommand\figurename{Supplementary Figure}
\renewcommand\tablename{Supplementary Table}
\newcommand\Scite[1]{[S\citealp{#1}]}
\newcommand\Scit[1]{S\citealp{#1}}

\makeatletter \renewcommand\@biblabel[1]{[S#1]} \makeatother

\textit{$\Gamma_e$ and $\Gamma_o$ for different corners of the charge stability diagram.} Here, we provide the details of the even- and odd-parity couplings presented in the main text.
For the corner in question, we first we find the center of the anti-crossing in $\varepsilon_{DL}-\varepsilon_{DR}$ plane.
We then obtain the excitation spectrum from the spectrum of the many-body Hamiltonian at the center of the anti-crossing:

\begin{align}
\delta E &= E^{\textrm{odd}}_{\textrm{gs}} - E^{\textrm{even}}_{\textrm{gs}},\\[3pt]
\delta E^{1}_{oe} &= E^{\textrm{odd}}_{1} - E^{\textrm{even}}_{\textrm{gs}},
\end{align}

where $E^{\textrm{even}(\textrm{odd})}_{\textrm{gs}}$ is the ground state energy in the even (odd) parity sector and $E^{\textrm{odd}}_{1}$ is the first excited state energy in the odd parity sector.
Using the excitation spectrum, we obtain the even- and odd-parity couplings:
\begin{align}
\Gamma_e &= \frac{\delta E^{1}_{oe} + \delta E}{2}, \\[3pt]
\Gamma_o &= \frac{\delta E^{1}_{oe} - \delta E}{2}. \\[3pt]
\end{align}
\begin{figure}[tbh]
\centering
{\includegraphics[width = 0.8\linewidth]{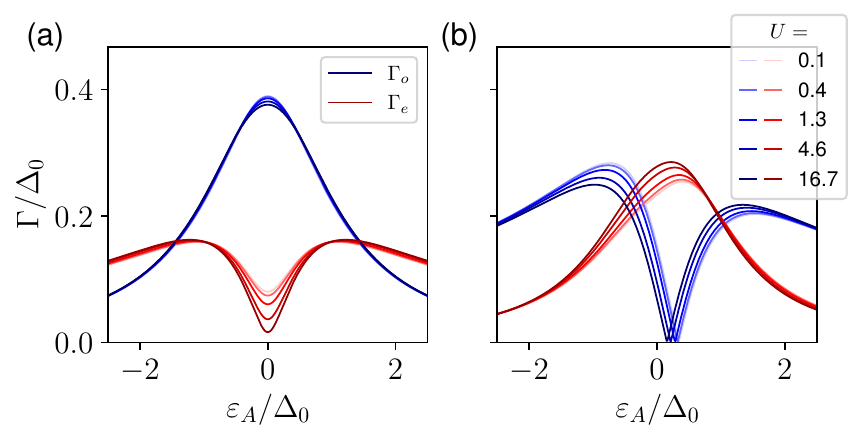}}
\caption{
$\Gamma_e$ (depicted by blue) and $\Gamma_o$ (depicted by orange) for (a) upper-left and (b) upper-right corner of the charge stability diagram for various values of $U_L = U_R = U$. Parameters used $\Delta_0 = 1, t=0.8, t_{so}=0.3t, E_z=1.5$.
}
\label{fig:Gamma_eo}
\end{figure}
In Fig.~\ref{fig:Gamma_eo}, we illustrate $\Gamma_e$ and $\Gamma_o$ for (a) the upper-left corner and (b) the upper-right corner of the charge stability diagram.
Varying the Coulomb repulsion strength on the outer dots, we change the corresponding BCS factors of YSR states.
As the Coulomb repulsion gets stronger, the BCS factors in the outer dots approaches to to their weak-coupling regime values, resulting in CAR and ECT like curves.
On the other hand, for weaker Coulomb repulsion, we observe that CAR and ECT terms get mixed for the upper-right corner, indicated by the shifting of the $\Gamma_e$ maxima and the $\Gamma_o$ minima.
Furthermore, the couplings of the upper-left corner only slightly change with respect to a variation in the Coulomb repulsion term, as expected from Eq.~\eqref{eqn:offdiagonal_gammas}.

\begin{figure}[t]
\centering
{\includegraphics[width = 0.8\linewidth]{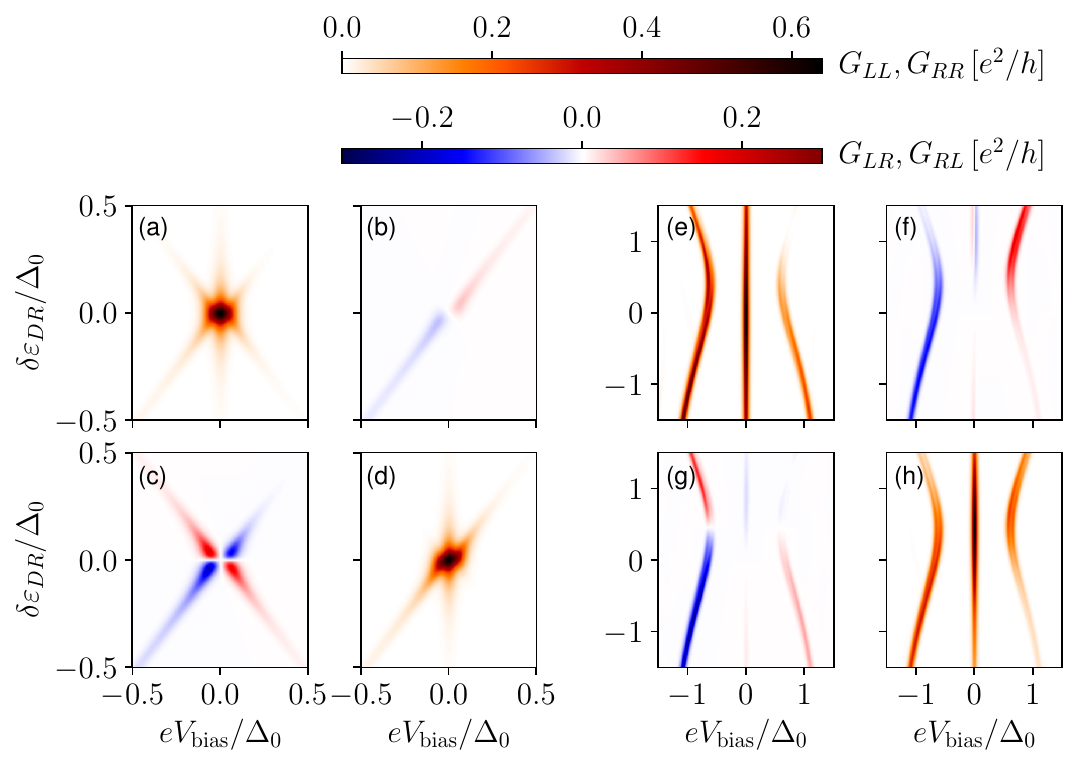}}
\caption{The elements of the conductance matrix in (a-d) the weak coupling ($t=0.25\Delta_0$) and (e-h) the strong coupling ($t=\Delta_0$) regime.}
\label{fig:conductance_matrix}
\end{figure}

\textit{Conductance matrix for weak and strong coupling regimes.} In the main text, we have shown how weak and strong coupling are reflected in the non-local conductances $G_{LR}$ and $G_{RL}$. In Fig.~\ref{fig:conductance_matrix} we show for completeness the full conductance matrix, including the local conductances $G_{LL}$ and $G_{RR}$.

In Fig.~\ref{fig:conductance_matrix}(a-d), we show the conductance matrix elements in the weak coupling regime.
As discussed in the main text, the nonlocal conductance $G_{LR}$ is close to zero in the weak coupling regime, while all the other conductance matrix elements exhibit large signals.
Moreover, in this regime, the reduced gap results in the lowering of the excited state energy, thereby merging with the zero-bias peak at the sweet spot.
This is reflected in the local conductance profiles shown in Fig.~\ref{fig:conductance_matrix}(a) and (d).

On the other hand, in the strong coupling regime, we observe an increased gap in the local conductance, shown in Fig.~\ref{fig:conductance_matrix}(e) and (h).
Simultaneously, the signal strength of the nonlocal conductance $G_{LR}$ also increases, making it distinguishable from the weak coupling case.

\begin{figure}[t]
\centering
{\includegraphics[width = 0.8\linewidth]{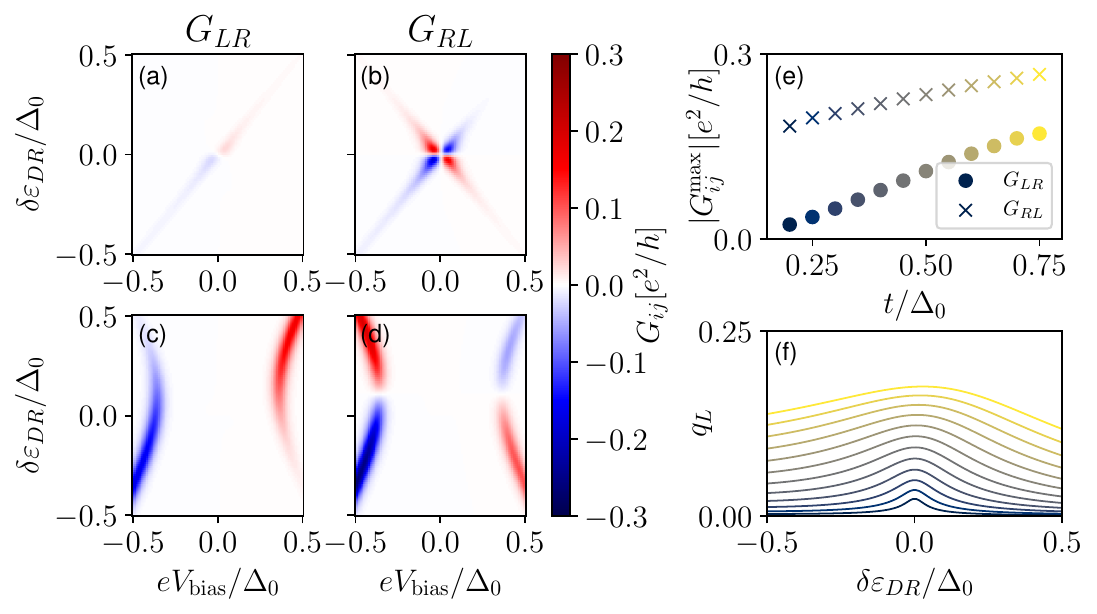}}
\caption{Spinless limit ($E_z=1000\Delta_0$) with no charging energy $U=0$ on the outer dots. Panels (a) and (b) show the nonlocal conductances in the weak coupling regime ($t=0.2\Delta_0$), whereas panels (c) and (d) are in the strong coupling regime ($t=0.75\Delta_0$). Similar to the results presented in the main text, panel (e) depicts how the strength of the nonlocal conductances vary with the dot-hybrid coupling $t$ and panel (f) shows the corresponding BCS charge of the excited state on the left quantum dot as we detune the right dot chemical potential away from the sweet spot by an amount $\delta\varepsilon_{DR}$. The qualitative features are similar to the case with finite charging energy presented in the main text.}
\label{fig:conductance_matrix_spinless}
\end{figure}

The increase in $G_{LR}$ with increasing coupling is also present even without Coulomb interaction and for a large Zeeman splitting. To demonstrate this, we compute the nonlocal conductances in the non-interactinf case in the nearly spinless limit, by setting $U=0$, and using a very high Zeeman splitting $E_z = 1000\Delta_0$ in the quantum dots.
In Fig.~\ref{fig:conductance_matrix_spinless}(a-b), we show
the behavior of nonlocal conductances $G_{LR}$ and $G_{RL}$ in weak coupling regime, $t=0.2\Delta_0$ ,as a function of the applied voltage bias and right dot chemical potential detuning.
Furthermore, in Fig.~\ref{fig:conductance_matrix_spinless}(c-d), we depict the nonlocal conductances in the strong coupling regime.
As we observe in Fig.~\ref{fig:conductance_matrix_spinless}(e), the nonlocal conductance $G_{LR}$ increases in signal strength as the dot-hybrid coupling $t$ is enhanced, a trend similar to what we observe in the interacting case.
This behavior, as explained in the main text, is a result of the finite BCS charge of the excited state induced on the left quantum dot as we vary the right chemical potential, which we illustrate in Fig.~\ref{fig:conductance_matrix_spinless}(f) for various $t$ values.
The fact that the same phenomenology also shows in a non-interacting system, justifies our use of the non-interacting BCS charge \cite{danon2020Nonlocal} to explain the behavior of the interacting system in the main text.

\end{document}